\documentclass[]{llncs}
\usepackage{graphicx}
\usepackage{xcolor}
\usepackage{multirow}
\usepackage{enumitem}
\usepackage{amssymb}
\usepackage{amsmath}
\usepackage{mathtools}
\usepackage{amsmath,systeme}
\usepackage{mathtools}
\usepackage{algorithm}

\begin{document}

\title{Highly Scalable Beaver Triple Generator from Additive-only Homomorphic Encryption}
\author{Huafei Zhu}
\institute{Agency for Science, Technology and Research, Singapore}
\maketitle
\begin{abstract}
In a convolution neural network, a composition of linear scalar product, non-linear activation function and maximum pooling computations are intensively invoked. As such, to design and implement privacy-preserving, high efficiency machine learning mechanisms, one highly demands a practical crypto tool for secure arithmetic computations. SPDZ, an interesting framework of secure multi-party computations is a promising technique deployed for industry-scale machine learning development if one is able to generate Beaver (multiplication) triple offline efficiently. This paper studies secure yet efficient Beaver triple generators leveraging privacy-preserving scalar product protocols which in turn can be constructed from additive-only homomorphic encryptions(AHEs). Different from the state-of-the-art solutions, where a party first splits her private input into a shared vector and then invokes an AHE to compute scalar product of the shared vectors managed by individual MPC server, we formalize Beaver triple generators in the context of 2-party shared scalar product protocol and then dispense the generated shares to MPC servers. As such, the protocol presented in this paper can be viewed as a dual construction of the state-of-the-art AHE based solutions. Furthermore, instead of applying the Paillier encryption as a basis of our previous constructions or inheriting from somewhat homomorphic encryptions, we propose an alternative construction of AHE from polynomial ring learning with error (RLWE) which results in an efficient implementation of Beaver triple generators. 
\end{abstract}

\section{Introduction}
In a convolution neural network (CNN for short), a party first performs a filtering computation leveraging the proposed kernel, then applies an activation function (say, relu) to the output of convolution layer and finally applies a sub-sampling/pooling strategy (say, max-pooling) to the output of the relu function. This procedure repeats several times (depending on the number of layers defined by the CNN structure) before a sequence of full connection procedures will be applied to the output of the convolution neural network. Suppose Alice who holds a private CNN image classification model, where each layer can be abstracted as a weight vector $w$ =$(w_1, \cdots, w_n)$ and a bias $b$, collaboratively makes a prediction on a private input $x$ = $(x_1, \cdots, x_n)$ from by the player Bob. A secure computation of CNN can thus be reduced to that of $(b, w) \cdot (1, x)^{T}$ for filtering and full connection computations together with a comparison protocol for the relu procedure. Since different techniques will be applied for implementing secure scalar product and secure comparison protocols, we will focus on the secure scalar product protocol throughout the paper and leave the development of secure yet highly scalable comparison protocol for the future research. We remark that a scalar product protocol can be implemented using different ways. One may choose homomorphic encryption based solutions, or differential based solutions or secure multi-party computations. Each of method has its own pros and cons. For example, while the efficiency of fully homomorphic encryptions has been improved dramatically in the recent years, the computational complexity of multiplications of ciphertexts is still a bottleneck. For a differential privacy based computation, the raw data is polluted. While a secure multi-party computation based solution in the framework of SPDZ provides an efficient way to compute arithmetic addition and multiplication, the generation of offline multiplication (Beaver) triples is challenging.     

Since the introduction of secure multi-party computation, numerous practical yet secure multi-party computation (MPC) protocols have been developed~(\cite{Yao86,Goldreichbook1,Goldreichbook2,Lindell201601} and references therein). The state-of-the-art solutions benefit us to construct highly scalable, high efficiency and secure multiplication triple generators leveraging the available technologies. The notable developments among them are ring-based zero-splitting~\cite{Dan2012,aby15,Dan2016,Lindell2016} and SPDZ-based solutions~\cite{CBC97,Ivan2012,Ivan2013,Keller16,Ivan2018,Smart1901,Smart2001}. Each of categories has its own pros and cons. For example, while a ring-based zero-splitting solution is efficient for 3-party computations, its scalability is problematic. On the other hand, while SPDZ provides high scalability, the efficiency of multiplication (or Beaver) triple generation is a bottleneck~\cite{Beaver91}. Note that a computation of arithmetic addition within the SPDZ framework is as efficient as that within the ring-based zero-splitting framework and also an arithmetic multiplication can be computed efficiently in the SPDZ framework if an auxiliary multiplication (Beaver) triple that is used to assist the multiplication is available among the MPC servers during the processing. That is, an efficient Beaver triple generator results in an efficient implementation of MPC trivially. An interesting research problem thus is $-$ \emph{how to construct highly scalable, high efficiency and secure multiplication (Beaver) triples for industry scale deployment?} 

\subsection{This work}
The Beaver triples deployed in the SPDZ framework are originally constructed from somewhat homomorphic encryptions~\cite{Ivan2012}. Since the computation cost of ciphertext multiplications is high if one leverages somewhat homomorphic encryptions, more and more researchers are considering alternative constructions such as oblivious-transfers~\cite{Keller16}, additively homomorphic encryptions~\cite{Keller18,AHE}, multiplicatively homomorphic encryption~\cite{MHE} and multiplicatively homomorphic key management system~\cite{HKM}, where a proof of equivalence between constructions leveraging multiplicatively homomorphic encryption and multiplicatively homomorphic key management system has been proposed. In this work, an efficient solution for generating Beaver triples starting from asymmetric oblivious scalar product protocol is proposed and analyzed. The notation of asymmetric oblivious scalar protocol was first introduced and formalized by Zhu and Bao in 2006~\cite{Zhu06A}. Asymmetric oblivious scalar protocol allows two parties Alice and Bob to collaboratively compute scalar-product obliviously so that at the end of the protocol, Alice learns $\sum_{i=0} ^l x_i y_i$ while Bob learns nothing, where Alice has an input vector $inp_A$ = $(1, x_1, \cdots, x_l)$ ($x_i \in F$, $1 \leq i \leq l$, $F$ is a finite field) and Bob has an input vector $inp_B$ = $(y_0, y_1, \cdots, y_l)$ ($y_i \in F$, $0 \leq i \leq l$). As discussed above, we know that an asymmetric oblivious scalar protocol can be viewed as an abstraction of privacy-preserving convolution neural network computation.

It has already shown that the notion of oblivious scalar protocol in essence, is a shared scalar product protocol and a construction of oblivious scalar protocol leveraging the Paillier's additively homomorphic encryption (~\cite{P99}) has been proposed~\cite{Zhu06B,Zhu06C}. Instead of directly applying the state-of-the-art Paillier's encryption as a basis presented in our previous constructions, we investigate an alternative construction from the polynomial ring learning with error (ring-LWE, or RWLE for short). As demonstrated in Section~4, the proposed additive-only homomorphic encryption scheme is more efficient compared with the additive-only Paillier's encryption, it follows that we are able to propose an efficient implementation of Beaver triple generators. 

Different from the state-of-the-art solutions proposed in~\cite{Keller18,AHE}, where a party (say, Alice) first splits her private input $\mathbf{x_A}$ to a share vector $\mathbf{x_A}$ = $(x_{A,1}, \cdots, x_{A, l})$ and then invokes an AHE to compute shares $s_A$ with Bob whose input is $\mathbf{x_B}$ and $\mathbf{x_B}$ = $(x_{B,1}, \cdots, x_{B, l})$ and output is $s_B$ such that $\mathbf{x_A} \cdot \mathbf{x_B}$ =$s_A + s_B$, we formalize Beaver triple generators in the context of 2-party shared scalar product protocol to get $s_A +s_B$ = $\mathbf{x_A \cdot x_B}$ and then Alice (Bob resp.,) dispenses her shares $\bf{s_A}$ = $(s_{A,1}, \cdots, s_{A, l})$ ($\bf{s_B}$ = $(s_{B,1}, \cdots, s_{B, l})$ resp,.) to MPC servers. As such, the protocol presented in this paper can be viewed as a dual construction of the state-of-the-art AHE based solutions.   

\subsection{Secure CNN computation within SPDZ}
We now provide a generic view to demonstrate why SPDZ is a promising framework for secure computation of CNN layered function $(b, w) \cdot (1, x)^{T}$. Let MPC$_1$, $\cdots$, MPC$_m$ be $m$-party computational servers and $[x]$ be a share of $x$ among the $m$-party. Assuming that MPC$_1$ (say, Alice) holds private data $x$ (for simplicity, we view parameters of CNN as private data of Alice) and MPC$_2$ (say, Bob) holds private data $y$. Suppose $m$ parties MPC$_1$, $\cdots$, MPC$_m$ wish to compute $[xy]$ collaboratively. MPC$_1$ first selects an auxiliary value $x_A \in F$ uniformly at random and then invokes an additive-only homomorphic encryption (AHE) to perform a secure 2-party computation with MPC$_2$ whose auxiliary value is $x_B \in F$ ($F$ is underlying finite field). Let $[s_A]$ ($[s_B]$ resp.,) be a share vector of $s_A$ ($s_B$ resp.,). Borrowing the notation from SPDZ, by $\rho$, we denote an opening of $[x] - [x_A]$ and by $\epsilon$, an opening of $[y] - [x_B]$. Here an opening refers to a procedure where all participants send their shares to the initiator (either Alice or Bob in our case) via established secure channels. Since a secure channel between two parties can be easily implemented assuming the existence of public key infrastructure (PKI), we simply assume that there is a secure channel between each pair of MPC servers throughout the paper. Given $\rho$ and $\epsilon$, each party can compute secret shares $[x y]$ of $xy$ as follows: $[x y]$ =  $(\rho + [x_A])$ $(\epsilon + [x_B])$ = $[x_A x_B]$ + $\epsilon [x_A]$ + $\rho [x_B]$ + $\rho \epsilon $ = $[s_A ]$ + $[s_B]$ + $\epsilon [x_A]$ + $\rho [x_B]$ + $\rho \epsilon $. As a result, if we are able to propose an efficient yet secure computation of $x_A x_B$ = $s_A +s_B $, the SPDZ provides a promising framework to compute $[xy]$ indeed.

\medskip

\subsubsection{The road-map} The rest of this paper is organized as follows: in Section~2, syntax, functionality and security definition of shared scalar product are proposed. An interesting additive-only homomorphic encryption from polynomial ring learning with error is constructed and analyzed in Section~3. We apply the developed additive-only homomorphic encryption to generate Beaver triples and construct a secret sharing mechanism and dispensing protocol in Section~4. We conclude our work in Section~5. 

\section{Syntax, functionality and security definition of shared scalar product}

\subsection{Syntax}
A shared scalar product protocol consists of the following two probabilistic polynomial time (PPT) Turing machines:
\begin{itemize}
\item On input system parameter $l$, a PPT Turing machine $A$ (say, Alice), chooses $l$ elements $ x_1, \cdots, x_l \in F$ uniformly at random (throughout the paper, we assume that $F$ =$Z_m ^*$, where $m$ is a large prime number). The input vector of Alice is denoted by $inp_A$=$(x_1, \cdots, x_l)$;

\item On input system parameter and $l$, a PPT Turing machine $B$ (say, Bob), chooses $l$ elements $y_1, \cdots, y_l \in Z_m$ uniformly at random. The input vector of Bob is denoted by $inp_B$=$(y_1, \cdots, y_l)$;

\item On inputs $inp_A$ and $inp_B$, Alice and Bob jointly compute the value $\sum_{i=1} ^l x_i y_i$ mod $m$;

\item The output of Alice is $s_A$ while Bob is $s_B$ such that $\sum_{i=1} ^l x_i y_i$~mod~$m$ = $s_A + s_B$.
\end{itemize}

\subsection{Functionality}
The functionality $\mathcal{F}_{SSP}$ of shared scalar product protocol (SSP) can be abstracted as follows:
\begin{itemize}
\item A player (say Alice) has her input vector $inp_A$=$(x_1, \cdots, x_l)$; Another player (say Bob) has his input vector $inp_B$=$(y_1, \cdots, y_l)$; Each participant sends the corresponding input set to $\mathcal{F}_{SSP}$ $-$ an imaginary trusted third party in the ideal world via a secure and private channel.

\item Upon receiving $inp_A$ and $inp_B$, $\mathcal{F}_{SSP}$ checks whether $x_i \in Z_m$ and $y_i \in Z_m$ ($1\leq i \leq l$). 

If the conditions are satisfied, then $\mathcal{F}_{SSP}$ computes $\sum_{i=1} ^l x_i y_i$ mod $m$;

If there exists a subset $\tilde{s_A}$ $\subset$ $inp_A$ such that each $x_i \in \tilde{s_A}$ but $x_i \notin Z_m$, then $\mathcal{F}_{SSP}$ chooses an element $x'_i \in_r Z_m$ and substitutes $x_i$ with $x'_i$. Similarly, if there exists a subset $\tilde{s_B}$ $\subset$ $inp_B$ such that each $y_i \in \tilde{s_B}$ but $y_i \notin Z_m$, then $\mathcal{F}_{SSP}$ chooses an element $y'_i \in_r Z_m$ and substitutes $y_i$ with $y'_i$. By $inp_A$=$(x_1, \cdots, x_l)$ (using the same notation of the input vector of Alice) we denote the valid input set of Alice which may be
modified by $\mathcal{F}_{SSP}$ and by $inp_B$=$(y_1, \cdots, y_l)$ (again using the same notation of the input vector of Bob), we denote the valid input set of Bob which may be modified of the original input values by $\mathcal{F}_{SSP}$. Once given the valid input sets $inp_A$ and $inp_B$, $\mathcal{F}_{SSP}$ computes $\sum_{i=} ^l x_i y_i$~mod~$m$. 

Finally, $\mathcal{F}_{SSP}$ sends $s_A$ to Alice via the secure and private channel while Bob learns $s_B$ such that $\sum_{i=1} ^l x_i y_i$~mod~$m$.

\item The output of Alice (Bob resp.) is $s_A$ ($s_B$ resp.) which is sent by $\mathcal{F}_{SSP}$ via the secure and private channel between them.
\end{itemize}

\begin{remark}
Notice that for semi-honest adversary, $\mathcal{F}_{SSP}$ does not check the input from Alice or Bob. That is, upon receiving $inp_A$ and $inp_B$, $\mathcal{F}_{SSP}$ assumes that both Alice and Bob follow the protocol. $\mathcal{F}_{SSP}$ simply selects $s_A$ and $s_B$ uniformly at random such that $s_A$ + $s_B$ =$\sum_{i=1} ^l x_i y_i$ mod $m$.
\end{remark}

\subsection{Security definition}
The security definition of shared scalar product protocols is defined in terms of the ideal-world vs. real-world framework. In this framework, we first consider an ideal model in which two dummy participants join in the imaginary trusted third party (TTP) ideal world. Note that the task of our construction is to remove such an imaginary TTP which does not exit in the real world. All performances are then computed via this trusted party. Next, we consider the real model in which a real two-party protocol is executed. A protocol in the real model is said to be secure with respect to certain adversarial behavior if the possible real execution with such an adversary can be simulated in the ideal model. That is, we want to show that there exists a polynomial time transform of adversarial behavior in the real conversation into corresponding adversarial behavior in the ideal model. We follow the security definitions of our previous work~\cite{Zhu06A,Zhu06B,Zhu06C}.

\begin{definition}
A shared scalar product protocol is secure against malicious (semi-honest resp.) Alice $A$, there exists a simulator $sim_A$ that plays the role of $A$ in the ideal world such that for any probabilistic polynomial time distinguisher $D$, the view of $D$ when it interacts with $A$ in real conversation is computationally indistinguishable from that when it interacts with $sim_A$ in the ideal world.
\end{definition}

\begin{definition}
A shared scalar product protocol is secure against malicious (semi-honest resp.) Bob $B$, if there exists a simulator $sim_B$ that plays the role of $B$ in the ideal world such that for any probabilistic polynomial time distinguisher $D$, the view of $D$ when it interacts with $B$ in real conversation is computationally indistinguishable from that when it interacts with $sim_B$ in the ideal world.
\end{definition}

\begin{definition}
A shared scalar product protocol is secure for any static probabilistic polynomial time (PPT) adversary if it is secure for any PPT Alice and any PPT Bob.
\end{definition}

\section{Additive-only homomorphic encryption}
Additively homomorphic encryption scheme can be inherited from somewhat homomorphic encryption scheme. The state-of-the-art somewhat homomorphic encryption is efficient if only additive property is deployed. Leveraging this idea, we construct our AHE from the RLWE assumption. 

\subsection{Additive-only homomorphic encryption based on ring-LWE}
Let $f(x)=x^n +1$ be a cyclotomic polynomial, i,e., the minimal polynomial of primitive roots of unity with $n=2^d$. Let $R$ = $\mathbb{Z}[x]/(f(x))$ and elements of the ring $R$ will be denoted in lowercase bold, e.g. $\mathbf{a} \in R$. The coefficient of an element in $R$ will be denoted by $a_i$ such that $\mathbf{a} = \sum_{i=0} ^{n-1} a_i x^i$. The infinite norm $||\mathbf{a}||$ is defined as max$_i |a_i|$, and the ratio of expansion factor of $R$ is defined as $\delta_R$ = max$\{||{\mathbf{a} \cdot \mathbf{b}||}/{||\mathbf{a}||\cdot ||\mathbf{b}||}: \mathbf{a}, \mathbf{b} \in R\}$. 

Let $q$ be an integer and by $\mathbb{Z}_q$ we denote the set of integers $(-q/2, q/2]$.
Notice that $\mathbb{Z}_q$ is simply considered as a set and thus the notion of
$\mathbb{Z}_q$ is different from that of $\mathbb{Z}/{q\mathbb{Z}}$. Let $R_q$ be the set
of polynomials in $R$ with coefficients in $\mathbb{Z}_q$. For $a \in \mathbb{Z}$, we denote $[a]_q$ be the unique integer in $\mathbb{Z}_q$ with $[a]_q$ = $a$~mod~$q$. For $a, q \in \mathbb{Z}$, we define the remainder modulo $q$ by $r_q(a) \in [0, q-1]$. For $\mathbf{a} \in R$, we denote $[\mathbf{a}]_q$ the element of $R$ obtained by applying $[\cdot]_q$ to all its coefficients. Similarly, for $x \in R$, we use [$x$] to denote rounding to the nearest integer, $\lceil x \rceil$ rounding up to the nearest integer and $\lfloor x \rfloor$ rounding down the nearest integer. 

\begin{definition}
Decision $RLWE_{d, q, \chi}$ problem: for security parameter $\lambda$, let $f(x)$ a cyclotomic polynomial with degree $def(f)$ = $\phi(m)$ depending on $\lambda$ and $R$ = $\mathbb{Z}[x]/(f(x))$ and $q$ = $q(\lambda)$. For a random $\bf{s}$ $\in$ $R_q$ and a distribution $\chi$ over $R$, by $A^{(q)} _{\bf{s},  \chi}$ we denoted a distribution obtained by choosing a uniformly random element $\bf{a}$ $\leftarrow$ $R_q$ and a noise term $\bf{e} \leftarrow \chi$ and outputting $[\bf{a}, \bf{a} \cdot \bf{s} + \bf{e}]$. The decision $RLWE_{d, q, \chi}$ problem aims to distinguish between distributions $A^{(q)}_{\bf{s},  \chi}$ and the uniform distribution $U(R^2 _q)$. The hardness of decision $RLWE_{d, q, \chi}$ problem assumes that there is no PPT distinguisher for  the decision $RLWE_{d, q, \chi}$ problem. 
\end{definition}

\begin{remark}
As noted in~\cite{Fan12}, the distribution $\chi$ in general is not as simple as just sampling coefficients according to the Gaussian distribution $D^n _{\mathbb{Z}, \sigma}$ ($\mu$ =0 and standard deviation $\sigma$). However, for polynomial $f(x)=x^n +1$ and $n=2^d$, we can indeed define $\chi$ as $D^n _{\mathbb{Z}, \sigma}$. Also, notice that we can assume that  $[\bf{a}, \bf{a} \cdot \bf{s} + \bf{e}]$ $\approx$ $U(R^2 _q)$ for $\bf{s} \in \chi$ chosen uniformly at random.
\end{remark}

\subsection{A new construction of polynomial-ring LWE encryption}
Motivated by the work of Fan and Vercauteren~\cite{Fan12}, an efficient additively homomorphic scheme is presented and analysed. Our polynomial ring LWE encryption consists of the following algorithms (\texttt{KeyGen}, \texttt{Enc}, \texttt{Dec} and \texttt{Eval}):

\begin{itemize}
\item the key generation algorithm \texttt{KeyGen}: on input a security parameter
$\lambda$, \texttt{KeyGen} samples $\mathbf{a} \in R_q$ and $\mathbf{s}$,  $\mathbf{e}$
$\leftarrow$ $\chi$; the public key $pk$ $\stackrel{def}{=}$ $([-\mathbf{a} \cdot
\mathbf{s} + t \mathbf{e} ]_q , \mathbf{a})$ and secret key $sk$ $\stackrel{def}{=}$
$\mathbf{s}$. The output of KeyGen is $(pk, sk)$.

\item the encryption algorithm \texttt{Enc}: on input a message $\mathbf{m}$ $\in$ $R_t$,
let $\mathbf{p_0}$ = $pk[0]$ and $\mathbf{p_1}$ = $pk[1]$, \texttt{Enc} samples
$\mathbf{u}$, $\mathbf{e_0}$, $\mathbf{e_1}$ $\leftarrow$ $\chi$, and returns $ ct$ =
$([\mathbf{p_0} \mathbf{u}$ + $t \mathbf{e_0}$ + $\mathbf{m}]_q, [\mathbf{p_1} \mathbf{u}$ + $t
\mathbf{e_1}]_q)$.

\item the decryption algorithm \texttt{Dec}: on input $ct$, let $\mathbf{c_0}$ =$ct[0]$,
$\mathbf{c_1}$ =$ct[1]$ and the message $\mathbf{m} \in \mathbb{Z}_t$ is computed from
$[[c_0 + c_1 \mathbf{s}]_q]_t$.

\item the evaluation algorithm \texttt{Eval}: on input two ciphertexts $ct_1$ and $ct_2$, \texttt{Eval} outputs a ciphertext $c$ such that \texttt{Dec}$c$ = \texttt{Dec}$c_1$  + \texttt{Dec}$c_2$.
\end{itemize}
The correctness is following from the lemma described below:
\begin{equation} \label{eq1}
\begin{split}
\mathbf{c_0} + \mathbf{c_1} \mathbf{s} & = \mathbf{p_0} \mathbf{u} + t \mathbf{e_0} +
\mathbf{m} +  q \mathbf{k_0}  + (\mathbf{p_1} \mathbf{u} + t \mathbf{e_1}) \mathbf{s} +
q \mathbf{k_1} \\ & = \mathbf{m} +( -\mathbf{a}\mathbf{s} + t \mathbf{e} + q \mathbf{k})
\mathbf{u} +  t \mathbf{e_0} + (\mathbf{a} \mathbf{u} + t \mathbf{e_1}) \mathbf{s} + q
(\mathbf{k_0} + \mathbf{k_1} \mathbf{s}) \\ & = \mathbf{m} + t (\mathbf{e}\mathbf{u} +
\mathbf{e_0} + \mathbf{e_1} \mathbf{s}) + q (\mathbf{k} \mathbf{u} + \mathbf{k_0} +
\mathbf{k_1}\mathbf{s})
\end{split}
\end{equation}
Since $[\mathbf{c_0} + \mathbf{c_1} \mathbf{s}]_q$ = $\mathbf{m} + t
(\mathbf{e}\mathbf{u} + \mathbf{e_0} + \mathbf{e_1} \mathbf{s})$, it follows that
$[[\mathbf{c_0} + \mathbf{c_1} \mathbf{s}]_q]_t $ = $\mathbf{m}$. The proof of security is extract the same as that presented in~\cite{Fan12} and thus it is omitted.  

\subsection{Additive and scalar properties}
Let $\mathbf{c}$ = ($\mathbf{ct[0]}$, $\mathbf{ct[1]}$) be an encryption of message $\mathbf{m} \in \mathbb{Z}_t$. Let$\mathbf{c'}$ = ($\mathbf{ct'[0]}$, $\mathbf{ct'[1]}$) be an encryption of message $\mathbf{m'} \in \mathbb{Z}_t$. It follows that $\mathbf{ct[0]}$ + $\mathbf{ct'[0]}$ = $\mathbf{p_0}\mathbf{u}$ + $t \mathbf{e_0} $ + $\mathbf{m}$ +  $q \mathbf{k_0}$ + $\mathbf{p_0}\mathbf{u'}$ + $t \mathbf{e'_0} $ + $\mathbf{m'}$ +  $q \mathbf{k'_0}$ = $\mathbf{p_0}(\mathbf{u} + \mathbf{u'})$ + $t(\mathbf{e_0} + \mathbf{e'_0})$ + $(\mathbf{m} + \mathbf{m'}) $ + $q (\mathbf{k_0} + \mathbf{k'_0})$ and $\mathbf{ct[1]} + \mathbf{ct'[1]}$ =$ \mathbf{p_1}( \mathbf{u} + \mathbf{u'})$ + $t (\mathbf{e_1} + \mathbf{e'_1})$ + $q (\mathbf{k_1} +\mathbf{k'_1})$. Applying the decryption procedure above, one gets addition $(\mathbf{m} +\mathbf{m'})$ from the aggregated ciphertext with the help of the secret key $\mathbf{s}$. This means that the proposed encryption scheme is additively homomorphic. 

\section{The construction and proof of security of Beaver triples}
In this section, we are able to propose a new Beaver triple generator and dispensing protocol leveraging the proposed additive-only homomorphic encryption. 

\subsection{Beaver triple functionality}
We write $[x]$ to mean that each party $P_i$ holds a random, additive sharing $x_i$ of $x$ such that $x$ = $x_1 + \cdots + x_n$, where $i=1, \cdots, n$. The values are stored in the dictionary Val defined in the functionality $\mathcal{F}_{\rm {Triple}} $~\cite{Keller16}. Please refer to the Table.1 for more details

\begin{table}
\label{table}
\noindent \framebox{\parbox[c]{12cm}{\center{\underline{The functionality of Beaver triple generator $\mathcal{F}_{\rm {BTG}} $}} 
\begin{itemize}
\item The functionality maintains a dictionary, Val to keep track of assigned value, where entry of Val lies in a fixed field. 
\item On input (Triple, id$_A$,id$_B$, id$_C$) from all parties, sample two random values $a$, $b$ $\leftarrow$ $F$, and set [ Val[id$_A$], Val[id$_B$], Val[id$_C$] ] $\leftarrow $ ($a$, $b$, $ab$).
\end{itemize}
}}
\medskip
\caption{Beaver triple functionality}
\end{table}

\subsection{The construction}
We propose an efficient implementation of SSP based on additive-only homomorphic encryption described above. We assume that public and secret key pair of AHE are generated by Alice. We consider a semi-honest adversary and our protocol is described as follows:
\begin{enumerate}
\item Let $x_A \in F$ be a random value and $c_A$ be a ciphertext AHE$(x_A)$. Alice sends $c_A$ to Bob;

\item Upon receiving $c_A$, Bob randomly selects a value $r_B \in F$ and then computes $c_B$ $\leftarrow$ $x_B \times c_A$ + AHE$(r_B)$;

\item Upon receiving $c_B$, Alice decrypts $c_B$ to get $s_A$ = $x_A x_B + r_B$. 

\item Alice output $s_A$ while Bob outputs $s_B$ =$-r_B$. 
\end{enumerate}
The correctness of protocol can be easily verified and thus it is omitted. The rest of our work is to show that the proposed scheme is secure against static semi-honest adversary.

\subsection{The proof of security}
\begin{theorem}
The proposed shared product protocol is secure against static and semi-honest adversary assuming that the underlying $\mathrm{AHE}$ is semantically secure.
\end{theorem}
\begin{proof}
Suppose Alice gets corrupted. We construct a simulator as follows:
\begin{itemize}
\item The simulator sim$_A$ invokes Alice to generate a pair of public and secret keys $(pk_A, sk_A)$. sim$_A$ is given $pk_A$ and $sk_A$ (in this paper, we are considering the static adversary). 

\item sim$_A$ then corrupts the corresponding dummy Alice in the ideal world, and gets to know the input $x_A$ and randomness $r_A$ used for the real world protocol execution. 

\item sim$_A$ sends $x_A$ to and gets $s_A$ from the functionality $\mathcal{F}_{\rm {BTG}}$ on behalf of the corrupted Alice. 
\end{itemize}
sim$_A$ then generates a random ciphertext $c_B$ on as a simulation of honest Bob's transcript. Notice that in the real world, the plaintext of $c_B$ is defined by $x_B x_A + r_B$ while in the simulation $c_B$ is an encryption of a random value. Since the underlying AHE is semantically secure, it follows that any probabilistic polynomial distinguisher cannot distinguish the simulated transcript from that of generated by the real world protocol.    

Now, we assume that Bob gets corrupted and construct a simulator as follows:
\begin{itemize}
\item Whenever Bob gets corrupted (in this paper, we are considering the static adversary), the simulator sim$_B$ corrupts the corresponding dummy Bob in the ideal world, and gets to know the input $x_B$ and the randomness $r_B$ assigned for the protocol execution. 

\item Upon receiving $c_A$, sim$_B$ gets $s_B$ from the functionality $\mathcal{F}_{\rm {BTG}}$ on behalf of the corrupted Bob. sim$_B$ then generates a ciphertext $c_B$ by computing $x_B c_A + \mathrm{AHE}(-s_B)$.  
\end{itemize}
A probabilistic polynomial time distinguisher's view on the simulated transcript is computationally indistinguishable from that generated in the real world. By combining sim$_A$ and sim$_B$, we know that the proposed scheme is secure against static semi-honest adversary. 
\end{proof}

\subsection{The dispense protocol}
Very recently, a block-chain based solution called TaaS (triple as a service) for dispensing Beaver triples leveraging commodity-based cryptography (CBC)~\cite{CBC97}, has been proposed. In the TaaS framework~\cite{TaaS}, the concept of ledger is introduced to make sure that the service providers do not reshare twice the same Beaver triple and every transaction has to be logged on a ledger. In this paper, we allows the data owner to control dispensations of Beaver triples and provide two solutions to dispense shares among MPC servers: one is public key based solution and another based on hybrid encryption scheme. 
\begin{itemize}
\item public-key solution: suppose that Alice gets $s_A$ and Bob gets $s_B$ such that $s_A + s_B$ = $x_A x_B$. Alice (Bob resp.,) can split $x_A$ and $s_A$ by randomly selecting $l$-tuple $(x_{A,2}, s_{A, 2})$, $\cdots$, $(x_{A,l}, s_{A, l})$. Alice then computes $x_{A, 1}$ from the equation $x_{A,1} + \cdots + x_{A, l}$ =$x_A$ and $s_{A,1}$ from the equation $s_{A,1} + \cdots + s_{A, l}$ =$s_A$. Each of shares is sent to Alice by an encryption of $x_{A,j}$ and $s_{A,j}$, where Alice holds private key of the encryption.

\item hybrid solution: alternative solution is to establish a secure channel between each pair of MPC servers and then encrypts shares using the shared session keys. Since these are standard crypto techniques, we omit the detail here.
\end{itemize}
We insist on the simple dispensing protocol above since a data owner control his/her randomness and hence his/her our data. 

\subsection{Experiment}
In this section, we provide experiment result of our additive-only homomorphic encryption scheme. The test environment is depicted below: the Python 3.8.1 works in the Window 10 with processors: Intel(R) Core(TM) i7-8665U CPU \@1.90GHz 2.11GHz; installed memory (RAM) 16.0GB (15.8 usable); and system type: 64-bit operating system, x64-based processor. The test parameters are described below:
\begin{itemize}
\item polynomial modulus degree: $n$ =$2^4$;

\item ciphertext modulus: $q$ = 140737488356903 (48-bit safe prime number);

\item plaintext modulus: $t$ =32843 (16-bit safe prime number);

\item polynomial modulus: poly\_mod = $x^n + 1$
\end{itemize}
For generating 1 million ciphertexts, the encryption algorithm costs about 300 seconds. To the best of our knowledge, this could be at least 2 to 3 orders efficiency gained compared with that leveraging the Paillier encryption scheme. As such, our design meets with the industrial deployments within the framework of numpy datatype int64. However, more work should be continuously investigated for very larger integer vectors beyond the scope of numpy datatype int64. This interesting task is opened to the research communities.
  
\section{Conclusion}
In this paper, we have proposed a new construction of multiplication triple generators leveraging asymmetric oblivious scalar product protocols. Then we have developed a new implementation of the scalar product protocol with the help of the additive-only homomorphic encryption schemes. An new implementation of additive homomorphic encryption scheme has been presented leveraging the ring-learning with error assumption. Our initial results have shown that the proposed scalar product protocol leveraging the proposed additive-only encryption scheme is more efficient than our previous work based on the Paillier's homomorphic encryption.


\begin{thebibliography}{2020}
\bibitem{aby15} D.Demmler et al: ABY - A Framework for Efficient Mixed-Protocol Secure Two-Party Computation. NDSS 2015.
\bibitem{Aono15} Y.Aono et al: Fast and Secure Linear Regression and Biometric Authentication with Security Update. IACR Cryptol. ePrint Arch. 2015: 692 (2015)
\bibitem{Aono17} Y.Aono et al: Efficient Key-Rotatable and Security-Updatable Homomorphic Encryption. SCC@AsiaCCS 2017: 35-42
\bibitem{Beaver91} D. Beaver: Efficient multiparty protocols using circuit randomization. Advances in Cryptology - CRYPTO '91
\bibitem{CBC97} D.Beaver: "Commodity-based cryptography (extended abstract)," in \emph{29\textsuperscript{th}} Annual ACM Symposium on Theory of Computing,  ACM Press, TX, USA, May 1997, pp.446-455.
\bibitem{Dan2012} D.Bogdanov et al: High-performance secure multi-party computation for data mining applications. Int. J. Inf. Sec. 11(6): 403-418 (2012)
\bibitem{Dan2016}D.Bogdanov et al.: Students and Taxes: a Privacy-Preserving Study Using Secure Computation, Proceedings on Privacy Enhancing Technologies, vol.3, 2016, pp.117-135.

\bibitem{Ivan2018}R. Cramer et al: SPD$_2^k$: Efficient MPC mod~$2^k$ for Dishonest Majority. IACR Cryptology ePrint Archive 2018: 482 (2018)


\bibitem{Ivan2012} I. Damgard, Valerio Pastro, Nigel P. Smart, Sarah Zakarias:
Multiparty Computation from Somewhat Homomorphic Encryption. CRYPTO 2012: 643-662.
\bibitem{Ivan2013}I. Damgard, Marcel Keller, Enrique Larraia, Valerio Pastro, Peter Scholl, Nigel P. Smart: Practical Covertly Secure MPC for Dishonest Majority - Or: Breaking the SPDZ Limits. ESORICS 2013: 1-18;

\bibitem{Fan12}J.Fan, F.Vercauteren: Somewhat Practical Fully Homomorphic Encryption. IACR Cryptol. ePrint Arch. 2012: 144 (2012)

\bibitem{Goldreichbook1} O.Goldreich, "The Foundations of Cryptography - Volume 1: Basic Techniques," Cambridge University Press, UK, 2001.
\bibitem{Goldreichbook2} O.Goldreich, "The Foundations of Cryptography, Volume 2 Basic Applications," Cambridge University Press, UK, 2004.

\bibitem{Keller16}M.Keller, E.Orsini, and P.Scholl, MASCOT: Faster Malicious Arithmetic Secure Computation with Oblivious Transfer, in 23rd ACM SIGSAC Conference on Computer and Communications Security, Vienna, Austria, October, 2016, pp. 830-842
\bibitem{Keller18} M.Keller, V.Pastro, D.Rotaru: Overdrive: Making SPDZ Great Again. EUROCRYPT (3) 2018: 158-189

\bibitem{Lindell201601} Y.Lindell, "How To Simulate It - A Tutorial on the Simulation Proof Technique," \emph{IACR Cryptol. ePrint Arch.} 216:46, 2016.
\bibitem{Lindell2016} T.Araki et al: High-Throughput Semi-Honest Secure Three-Party Computation with an Honest Majority. ACM Conference on Computer and Communications Security 2016: 805-817

\bibitem{Smart2001} E.Orsini, Nigel P. Smart, Frederik Vercauteren:
Overdrive2k: Efficient Secure MPC over $\mathbb {Z}_{2^k}$ from Somewhat Homomorphic Encryption. CT-RSA 2020: 254-283
\bibitem{P99} Pascal Paillier: Public-Key Cryptosystems Based on Composite Degree Residuosity Classes. Proc. of EUROCRYPT 1999: 223-238, Springer Verlag.
\bibitem{AHE} D. Rathee \emph{et al.}, "Improved Multiplication Triple Generation over Rings via RLWE-Based AHE," in \emph{18\textsuperscript{th}} International Conference, Fuzhou, China, October 2019, pp.347-359.
\bibitem{Smart1901} N.Smart, Titouan Tanguy: TaaS: Commodity MPC via Triples-as-a-Service. CCSW@CCS 2019: 105-116

\bibitem{TaaS} N.Smart \emph{et al.}, "TaaS: Commodity MPC via Triples-as-a-Service." In ACM SIGSAC Conference on Cloud Computing Security Workshop, London, UK, Nov 2019, pp.105-116.

\bibitem{Yao86} A.Yao: How to Generate and Exchange Secrets (Extended Abstract) FOCS 1986: 162 - 167.
\bibitem{Zhu06A} H.Zhu, F.Bao: Oblivious Scalar-Product Protocols. ACISP 2006: 313-323
\bibitem{Zhu06B} H.Zhu et.al: More on Shared-Scalar-Product Protocols. ISPEC 2006: 142-152
\bibitem{Zhu06C} H.Zhu et.al: Privacy-Preserving Shared-Additive-Inverse Protocols and Their Applications. SEC 2006: 340-350.
\bibitem{MHE} H.Zhu \emph{et al.}, "Privacy-Preserving Weighted Federated Learning Within the Secret Sharing Framework", IEEE Access vol.8, 2020, pp.198275 -198284.
\bibitem{HKM} H.Zhu, A Lightweight, Anonymous and Confidential Genomic Computing for Industrial Scale Deployment. CoRR abs/2110.01390 (2021)
\end{thebibliography}
\end{document}